\documentclass[9pt,twocolumn,twoside]{pnas-new}
\usepackage{aas_macros} % used by macros in .bib files.
\usepackage{units}
\usepackage{bm}% bold math
\usepackage{amssymb}
\usepackage{soul} %strike out the text
\usepackage{verbatim}
\usepackage{xcolor,afterpage}
\usepackage{subfigure}
\usepackage{xspace}
\usepackage{pnasresearcharticle}
\newcommand{\ie}{i.e.,\@\xspace}
\newcommand{\pred}{\mathrm{pred}}
\newcommand{\true}{\mathrm{true}}
\newcommand{\Mpch}{\,h^{-1}\mathrm{Mpc}}
\newcommand{\hMpc}{\,h\mathrm{Mpc^{-1}}}
\newcommand{\real}{\mathbb{R}}
\newcommand{\As}{{A_\mathrm{s}}}
\newcommand{\vk}{{\bm k}}

\newcommand{\vr}{{\bm r}}

\newcommand{\vq}{{\bm q}}

\newcommand{\vPsi}{\mathbf{\Psi}}
\renewcommand{\d}{\mathrm{d}}
\newcommand{\acr}[1]{\textsc{#1}}
\newcommand{\dm}{\acr{D$^3$M}\@\xspace}

\templatetype{pnasresearcharticle} 
\setboolean{displaywatermark}{false}

\title{Learning to Predict the Cosmological Structure Formation}

\author[a,b,c,1]{Siyu He}
\author[d,e,f]{Yin Li}
\author[d,e]{Yu Feng}
\author[c,e,d,a,b,1]{Shirley Ho}
\author[g]{Siamak Ravanbakhsh}
\author[c]{Wei Chen}
\author[h]{Barnab\'as P\'oczos}

\affil[a]{Physics Department, Carnegie Mellon University, Pittsburgh PA 15213}
\affil[b]{McWilliams Center for Cosmology, Carnegie Mellon University, Pittsburgh, PA 15213, USA}
\affil[c]{Center for Computational Astrophysics, Flatiron Institute, New York, NY 10010}
\affil[d]{Berkeley Center for Cosmological Physics, University of California, Berkeley, CA 94720, USA}
\affil[e]{Lawrence Berkeley National Laboratory, Berkeley, CA 94720, USA}
\affil[f]{Kavli Institute for the Physics and Mathematics of the Universe, University of Tokyo Institutes for Advanced Study, The University of Tokyo, Chiba 277-8583, Japan}
\affil[g]{Computer Science Department, University of British Columbia, Vancouver, BC V6T1Z4, Canada}
\affil[h]{Machine Learning Department, Carnegie Mellon University, Pittsburgh PA 15213}
\leadauthor{He} 

\significancestatement{
To understand the evolution of the Universe requires a concerted effort of accurate observation of the sky and fast prediction of structures in the Universe. N-body simulation is an effective approach to predicting structure formation of the Universe, though computationally expensive.
Here we build a deep neural network to predict structure formation of the Universe.
It outperforms the traditional fast analytical approximation, and accurately extrapolates far beyond its training data.
Our study proves that deep learning is
an accurate alternative to traditional way of generating approximate cosmological simulations.
Our study also used deep learning to generate complex 3D simulations in cosmology. This suggests deep learning can provide a powerful alternative to traditional numerical simulations in cosmology.} 
\authorcontributions{S. He, Y.L., S. Ho, and B.P. designed research; S. He, Y.L., Y.F., and S. Ho performed research; S. He, Y.L., S.R., and B.P. contributed new reagents/analytic tools; S. He, Y.L., Y.F., and W.C. analyzed data; and S. He, Y.L., Y.F., S. Ho, S.R., and W.C. wrote the paper.}

\authordeclaration{The authors declare no conflict of interest. \\ \\Data deposition: The source code of our implementation is available at \url{https://github.com/siyucosmo/ML-Recon}. The code to generate the training data is available at \url{https://github.com/rainwoodman/fastpm}. \\ \\ This article contains supporting information online at \url{www.pnas.org/lookup/suppl/doi:10.1073/pnas.1821458116/-/DCSupplemental}.}

\correspondingauthor{\textsuperscript{1}To whom correspondence may be addressed. Email: shirleyho@flatironinstitute.org or siyuh@andrew.cmu.edu.}


\begin{abstract}
Matter evolved under influence of gravity from minuscule density fluctuations.
Non-perturbative structure formed hierarchically over all scales, and developed non-Gaussian features in the Universe, known as the Cosmic Web.  
To fully understand the structure formation of the Universe is one of the holy grails of modern astrophysics.
Astrophysicists survey large volumes of the Universe 
and 
employ a large ensemble of computer simulations to compare with the observed data in order to extract the full information of our own Universe.
However, to evolve billions of particles over billions of years even with the simplest physics is a daunting task.
We build a deep neural network, the Deep Density Displacement Model (hereafter \dm), which learns from a set of pre-run numerical simulations, to predict the non-linear large scale structure of the Universe with Zel'dovich Approximation (hereafter ZA), an analytical approximation based on perturbation theory, as the input.  
Our extensive analysis, demonstrates that \dm outperforms the second order perturbation theory (hereafter 2LPT), the commonly used fast approximate simulation method, in predicting cosmic structure in the non-linear regime.  
We also show that \dm is able to accurately extrapolate far beyond its training data, and predict structure formation for significantly different cosmological parameters.
Our study proves that deep learning is 
a practical and accurate alternative to approximate 3D simulations of the gravitational structure formation of the Universe.
\end{abstract}

\doi{\url{https://doi.org/10.1073/pnas.1821458116}}

\begin{document}

\maketitle
\thispagestyle{firststyle}
\ifthenelse{\boolean{shortarticle}}{\ifthenelse{\boolean{singlecolumn}}{\abscontentformatted}{\abscontent}}{}

\dropcap{A}strophysicists require a large amount of simulations to extract the information from observations \cite{2dF, SDSS, 6df, GAMA, VIPERS, LSST, Euclid, WFIRST}. At its core, modeling structure formation of the Universe is a computationally challenging task; it involves evolving billions of particles with the correct physical model over a large volume over billions of years \cite{1998NewA....3..687M,2001NewA....6...79S,2002JApA...23..185B}.
To simplify this task, we either simulate a large volume with simpler physics or a smaller volume with more complex physics.
In order to produce the cosmic web \cite{CosmicWeb} in large volume, we select gravity, the most important component of the theory, to simulate at large scales. A gravity-only $N$-body simulation is the most popular; and effective numerical method to predict the full 6D phase space distribution of a large number of massive particles whose position and velocity evolve over time in the Universe \cite{Davis1985}. Nonetheless, $N$-body simulations are relatively computationally expensive, thus making the comparison of the $N$-body simulated large-scale structure (of different underlying cosmological parameters) with the observed Universe a challenging task. We propose to use a deep model that predicts the structure formation as an alternative to $N$-body simulations. 

Deep learning~\cite{lecun2015deep} is a fast growing branch of machine learning
where recent advances have lead to models that reach and sometimes exceed human performance across diverse areas,
from analysis and synthesis of images~\cite{huang2017densely,karras2017progressive,gulrajani2017improved}, sound~\cite{van2016wavenet,amodei2016deep}, text~\cite{hu2017toward,vaswani2017attention} and videos~\cite{denton2018stochastic,donahue2015long} to complex control and planning tasks as they appear in robotics and game-play~\cite{silver2016mastering,mnih2015human,levine2016end}. This new paradigm is also significantly impacting a variety of domains in the sciences, from biology~\cite{ching2018opportunities, alipanahi2015predicting} to chemistry~\cite{segler2018planning,gilmer2017neural} and physics~\cite{carleo2017solving,adam2015higgs}. In particular, in astronomy and cosmology, a growing number of recent studies are using deep learning for a variety of tasks, ranging from analysis of cosmic microwave background~\cite{he2018analysis,perraudin2018deepsphere,caldeira2018deepcmb}, large-scale structure~\cite{Ravanbaksh,mathuriya2018cosmoflow}, and gravitational lensing effects~\cite{Hezaveh,lanusse_lensing_generation} to classification of different light sources~\cite{kennamer18a,kim2016star,lochner2016photometric}.

\begin{figure}
\centering
\includegraphics[width=0.98\columnwidth]{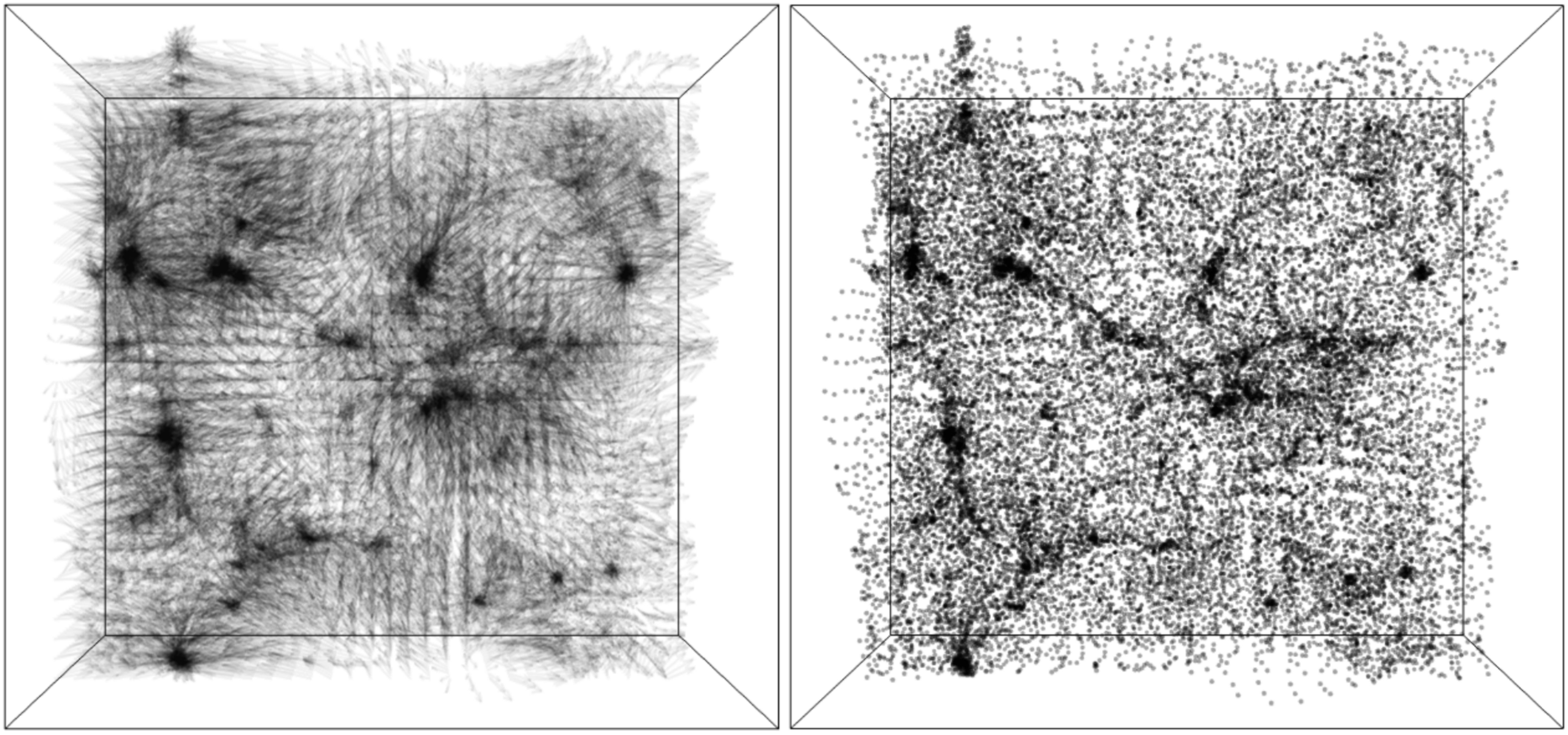}
\caption{(left) The displacement vector-field and (right) the resulting density field produced by \dm. The vectors in the left figure are uniformly scaled down for better visualization.}
\label{fig:d3m}
\end{figure}

The ability of these models to learn complex functions 
has motivated many to use them to understand the physics of interacting objects leveraging image, video and relational data
\cite{battaglia2013simulation,battaglia2016interaction,mottaghi2016newtonian,chang2016compositional,wu2015galileo,wu2016physics,watters2017visual,lerer2016learning,agrawal2016learning,fragkiadaki2015learning,tompson2016accelerating}.
However, modeling the dynamics of billions of particles in N-body simulations poses a distinct challenge.

In this paper we show that a variation on the architecture of a well-known deep learning model~\cite{Unet}, can efficiently transform the first order approximations of the displacement field and approximate the exact solutions, thereby producing accurate estimates of the large-scale structure. Our key objective is to prove that this approach is an accurate and computationally efficient alternative to expensive cosmological simulations, and to this end we provide an extensive analysis of the results in the following section.

The outcome of a typical N-body simulation depends on both the initial conditions and 
on cosmological parameters which affect the evolution equations. A striking discovery is that \dm, 
trained using a single set of cosmological parameters generalizes to new sets of significantly different parameters, minimizing the need for training data on diverse range of cosmological parameters. 

\section*{Setup}
We build a deep neural network, \dm, with similar input and output of an
$N$-body simulation.  
The input to our \dm  is the displacement field from ZA \cite{Zel70}. 
A displacement vector is the difference of a particle position at target redshift $z = 0$, \ie the present time, and its Lagrangian position on a uniform grid.
ZA evolves the particles on linear trajectories along their initial displacements.
It is accurate when the displacement is small, therefore
ZA is frequently used to construct the initial conditions of $N$-body simulations~\cite{MWhiteZA14}.
As for the ground truth, the target displacement field is produced using FastPM~\cite{FastPM}, 
a recent approximate N-body simulation scheme that is based on a particle-mesh (PM) solver. 
FastPM quickly approaches a full N-body simulation with high accuracy and provides a viable alternative to direct N-body simulations for the purpose of our study. 

A significantly faster approximation of N-body simulations is produced by second-order Lagrangian perturbation theory (hereafter 2LPT), which bends each particle's trajectory with a quadratic correction~\cite{Buchert94}. 
2LPT is used in many cosmological analyses to generate a large number of cosmological simulations for comparison of astronomical dataset against the physical model~\cite{JascheWandelt13, Kitaura13} or to compute the covariance of the dataset~\cite{BOSS, eBOSS, DESI}.
We regard 2LPT as an effective way to efficiently generate a relatively accurate description of the large-scale structure and therefore we select 2LPT as the reference model for comparison with \dm.

We generate 10,000 pairs of ZA approximations as input and accurate FastPM approximations as target.
We use simulations of $32^3$ $N$-body particles in a volume of $128 \Mpch$ (600 million light years, where $h=0.7$ is the Hubble parameter). 
The particles have a mean separation of $4 \Mpch$ per dimension. 

\begin{figure*}
\centering
\includegraphics[width=1.6\columnwidth]{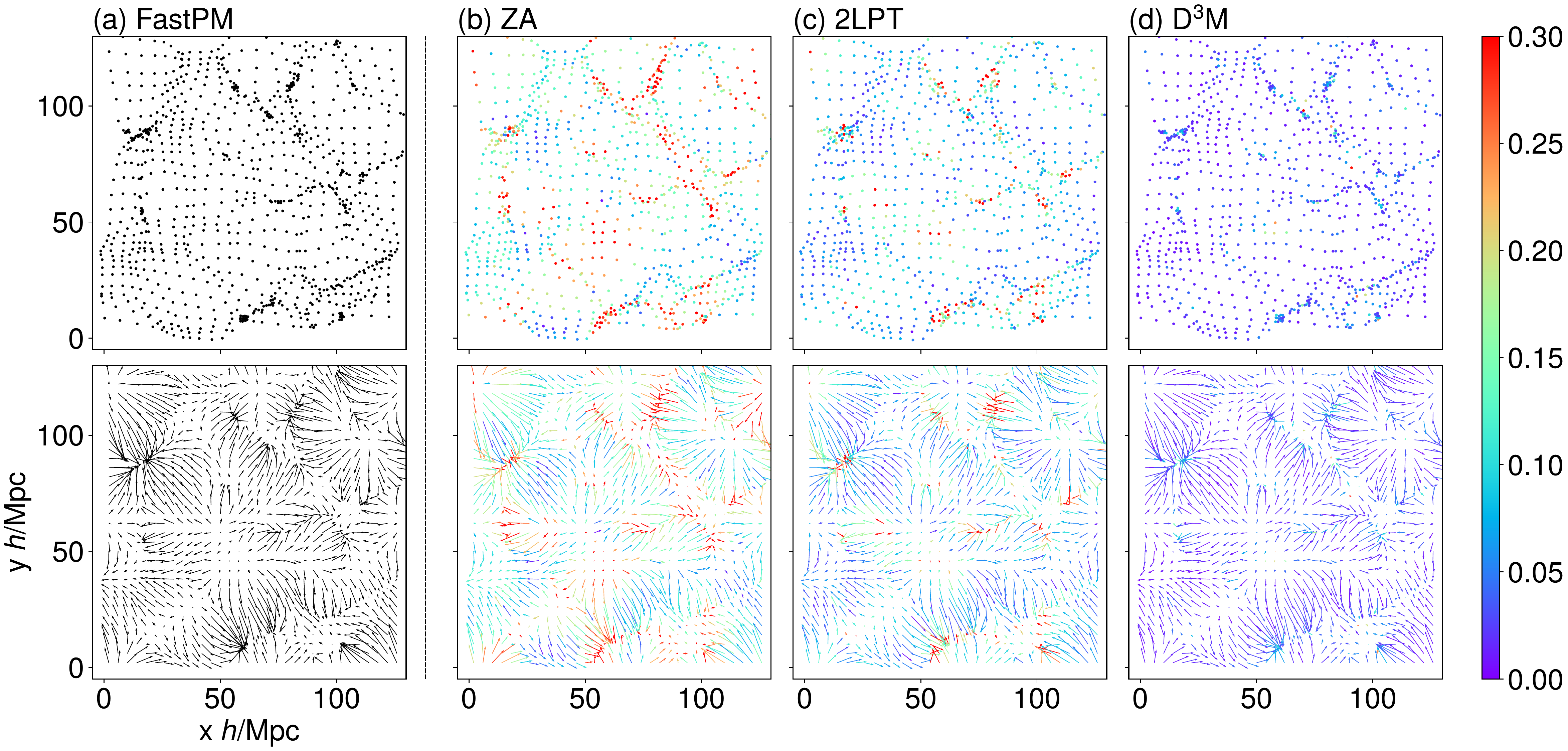}
\caption{
The columns show 2-D slices of full particle distribution (top) and displacement vector (bottom) by various models, from left to right:\\
%the color of the plot indicates the relative difference between the target location (displacement vector) and predicted distributions by various methods (b-d).
%while the columns to its right are relative differences between the prediction (by various models) and the ground truth. From left column to the right, we have \\
\textbf{(a)} FastPM: the target ground truth, a recent approximate N-body simulation scheme that is based on a particle-mesh (PM) solver ;\\
\textbf{(b)} Zel'dovich approximation (ZA): a simple linear model that evolves particle along the initial velocity vector; \\ 
\textbf{(c)} Second order Lagrangian perturbation theory (2LPT): a commonly used analytical approximatation;\\
\textbf{(d)} Deep learning model (\dm) as presented in this work. \\
While FastPM (A) served as our ground truth, B–D include color for the points or vectors. 
The color indicates the relative difference $(q_{model}-q_{target})/ q_{target}$ between the target (a) location or displacement vector and predicted distributions by various methods (b-d).
The error-bar shows denser regions have a higher error for all methods.
which suggests that it is harder to predict highly non-linear region correctly for all models: \dm, 2LPT and ZA.
Our model \dm has smallest differences between predictions and ground truth among the above models (b)-(d). 
}
\label{fig:error_vectors}
\end{figure*}

An important choice in our approach is training with displacement field rather than density field.
Displacement field $\mathbf{\Psi}$ and density field $\rho$ are two ways of describing the same distribution of particles. And an equivalent way to describe density field is the over-density field, defined as $\delta=\rho / \bar\rho - 1$, with $\bar\rho$ denoting the mean density. The displacement field and over-density field are related by eq.~\ref{eq:DenAndDis}.
\begin{equation}\label{eq:DenAndDis}
    \begin{split}
    \mathbf{x} &= \mathbf{\Psi}(\mathbf{q}) + \mathbf{q} \\
    \delta(\mathbf{x}) &= \int d^3 q \delta_D(\mathbf{x}-\mathbf{q}-\mathbf{\Psi}(\mathbf{q}))-1
    \end{split}
\end{equation}

%and they are related by the law of mass conservation.
%\begin{equation}
%\mathbf{\nabla} \cdot \mathbf{\Psi} = -\delta,
%\end{equation}
%The mean value of the density field is $\bar\rho$, relative to which the density fluctuation can be described by the over-density field, $\delta=\rho / \bar\rho - 1$.
%, with $\bar\rho$ denoting the uniform density. 

When the displacement field is small and has zero curl, the choice of over-density vs displacement field for the output of the model is irrelevant, as there is a bijective map between 
these two representations, described by the equation:
\begin{equation}
    \mathbf{\Psi} = \int \frac{d^3k}{(2\pi)^3}e^{i\mathbf{k} \cdot \mathbf{q}} \frac{i\mathbf{k}}{k^2}\delta(\mathbf{k})
\end{equation}
However as the displacements grow into the non-linear regime of structure formation, different displacement fields can produce identical density fields \cite[e.g.][]{FengSurface18}. 
Therefore, providing the model with the target displacement field during the training eliminates the ambiguity associated with the density field. Our inability to produce comparable results when using the density field as our input and target attests that relevant information resides in the displacement field (See SI Appendix, Fig. S1) . 

\section*{Results and Analysis}
Figure \ref{fig:d3m} shows the displacement vector field  as predicted by \dm (left) and the associated point-cloud representation of the structure formation (right). It is possible to identify structures such as clusters, filaments and voids in this point-cloud representation. 
We proceed to compare the accuracy of \dm and 2LPT compared with ground truth.

\subsection*{Point-Wise Comparison}
Let $\vPsi \in \real^{d \times d \times d \times 3}$ denote the displacement field, where $d$ is the number of spatial resolution elements in each dimension ($d=32$). A natural measure of error is the relative error $|\hat{\vPsi} - \vPsi_{t}| / |\vPsi_t|$, where $\vPsi_t$ is the true displacement field (FastPM) 
and $\hat{\vPsi}$ is the prediction from 2LPT or \dm. 
Figure~\ref{fig:error_vectors} compares this error for different approximations in a 2-D slice of a single simulation.
We observe that \dm predictions are very close to the ground truth, with a maximum relative error of 1.10 over all 1000 simulations. For 2LPT this number is significantly higher at 4.23.
%Since a few large error vectors can dominate the $L^2$ norm, we also measure the relative displacement error
%$|\hat{\vPsi} - \vPsi_{t}| / |\vPsi_t|$.
%Using this measure we observe the same pattern: \dm is almost 3 times more accurate than 2LPT.
In average, the result of \dm comes with a 2.8\% relative error while for 2LPT it equals 9.3\%.

\subsection*{2-Point Correlation Comparison}
As suggested by Figure~\ref{fig:error_vectors} the denser regions seem to have a higher error for all methods -- that is, more non-linearity in  structure formation creates larger errors for both \dm and 2LPT. The dependence of error on scale is computed with 2-point and 3-point correlation analysis. 

Cosmologists often employ compressed summary statistics of the density field in their studies. The most widely used of these statistics are the 2-point correlation function (2PCF) $\xi(r)$ and its Fourier transform, the power spectrum $P_{\delta \delta}(k)$:
\begin{equation}
    \begin{split}
    \xi(|\vr|) &= \langle \delta_A(\vr')\delta_B(\vr'+\vr) \rangle, \\
    P_{\delta \delta}(|\vk|) &= \int
    %\frac{d^3 \mathbf{r}}{(2\pi)^3}
    \!\d^3 \vr\;
    \xi(r)e^{i\vk \cdot \vr},
    \end{split}
\end{equation}
where the ensemble average $\langle\,\rangle$ is taken over all
possible realizations of the Universe.
%Assuming isotropy of the Universe,
Our Universe is observed to be both homogeneous and isotropic on large scales, i.e.\ without any special location or direction. This allows one to drop the dependencies on $\vr'$ and on the direction of $\vr$, leaving only the amplitude $|\vr|$ in the final definition of $\xi(r)$.
In the second equation, $P_{\delta \delta}(k)$ is simply the Fourier transform of $\xi(r)$, and captures the dispersion of the plane wave amplitudes at different scales in the Fourier space.
$\vk$ is the 3D wavevector of the plane wave, and its amplitude $k$ (the wavenumber) is related to the wavelength $\lambda$ by $k = 2\pi / \lambda$.
Due to isotropy of the Universe, we drop the vector form of $\vr$ and $\vk$.
%and in what follows we use $\xi(r)$ and $P(k)$.

Because FastPM, 2LPT and \dm take the displacement field as input and output, we also study the two-point statistics for the displacement field. The displacement power spectrum is defined as:
\begin{eqnarray}
P_{\boldsymbol{\Psi\Psi}}(k)=\langle \Psi_x(k)\Psi_x^*(k) \rangle+ \langle \Psi_y(k)\Psi_y^*(k) \rangle + \langle \Psi_z(k)\Psi_z^*(k)\rangle
\end{eqnarray}

\begin{figure*}[htbp]
\centering
\includegraphics[width=2\columnwidth]{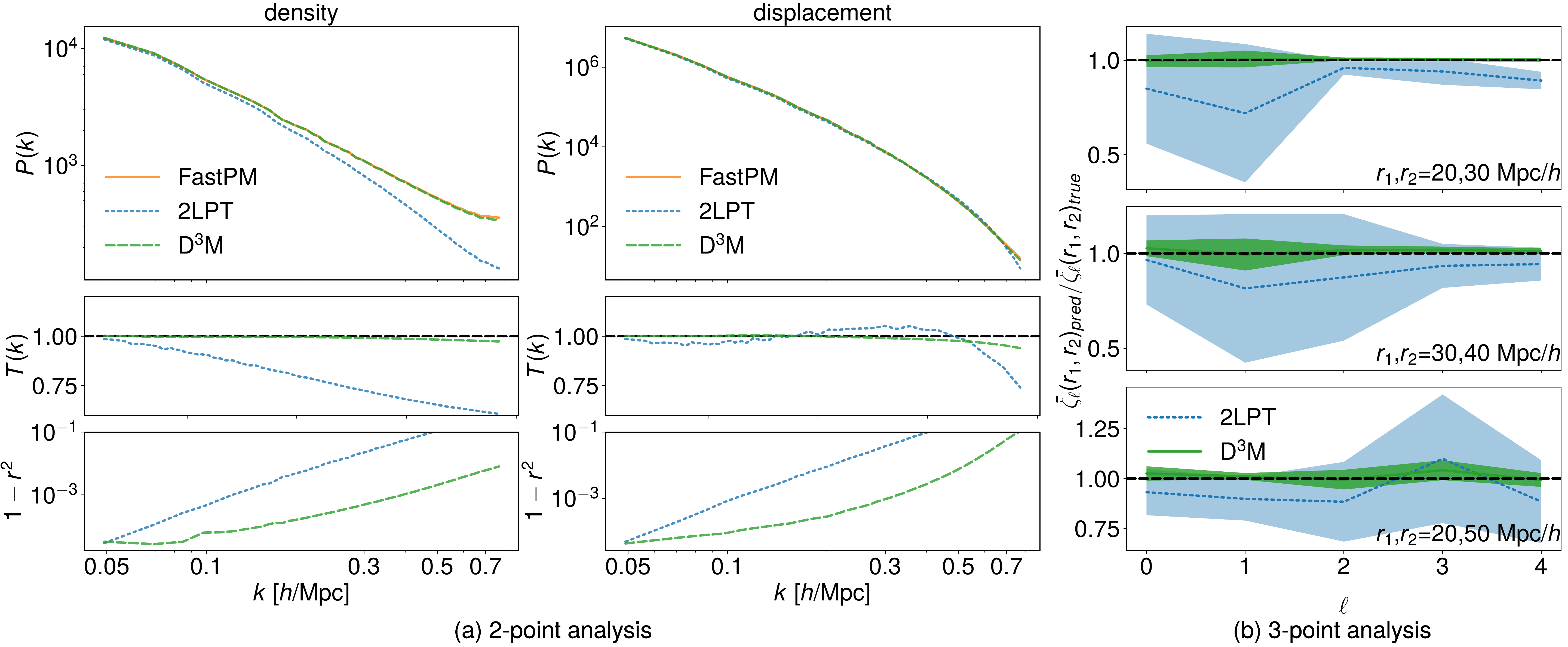}
\label{fig:3PCF}

\caption{(\textbf{a}) From top to bottom: (top) displacement and density power-spectrum of FastPM (orange), 2LPT (blue), and \dm (green); (middle) transfer function  -- \ie the square root of the ratio of the predicted power-spectrum to the ground truth; (bottom) 1-$r^2$ where $r$ is the correlation coefficient between the predicted fields and the true fields. Results are the averaged values of 1000 test simulations. The transfer function and correlation coefficient of the \dm predictions is nearly perfect from large to intermediate scales and outperforms our benchmark 2LPT significantly. \\
(\textbf{b}) 
The ratios of the multipole coefficients ($\zeta_l(r_1, r_2)$) (to the target) of the two 3-point correlation functions for several triangle configurations. The results are averaged over 10 test simulations. The error-bars (padded regions) are the standard deviations derived from 10 test simulations. The ratio shows the 3-point correlation function of \dm is closer than 2LPT to our target FastPM with lower variance.
}
\label{fig:test}
\end{figure*}

%YL: I moved this sentence to the previous paragraph
%The power spectrum captures the variance of the field of interest at different scales in the Fourier space. We consider cross power spectrum and covariance when we look at two different fields. In addition,
We focus on the Fourier-space representation of the 2-point correlation. 
Because the matter and the displacement power spectrum take the same form, in what follows we drop the subscript for matter and displacement field and use $P(k)$ to stand for both matter and displacement power spectrum. 
We employ the transfer function $T(k)$ and the correlation coefficient $r(k)$ as metrics to quantify the model performance against the ground truth (FastPM) in the 2-point correlation. We define the transfer function $T(k)$ as the square root of the ratio of two power spectra,
\begin{equation}
\label{transfer}
T(k) = \sqrt{\frac{P_\pred(k)}{P_\true(k)}},
\end{equation}
where $P_\pred(k)$ is the density or displacement power spectrum as predicted by 2LPT or
\dm, and analogously $P_\true(k)$ is the ground truth predicted by FastPM. The correlation coefficient r(k) is a form of normalized cross power spectrum,
\begin{equation}
\label{corrcoeff}
r(k) = \frac{P_{\pred\times\true}(k)}{\sqrt{P_\pred(k)P_\true(k)}},
\end{equation}
where $P_{\pred \times \true}(k)$ is the cross power spectrum between 2LPT or
\dm predictions and the ground truth (FastPM) simulation result.
The transfer function captures the discrepancy between amplitudes, while the correlation coefficient can indicate the discrepancy between phases 
as functions of scales. For a perfectly accurate prediction, $T(k)$ and $r(k)$ are both 1. In particular, $1-r^2$ describes stochasticity, the fraction of the variance
in the prediction that cannot be explained by the true model.

Figures~\ref{fig:test}(a) shows the average power spectrum, transfer function $T(k)$  and
stochasticity $1-r^2(k)$ of the displacement field  and the density field over 1000 simulations. 
The transfer function of density from 2LPT predictions is 2\% smaller than that of FastPM on large scales ($k \approx 0.05 \hMpc$). This is expected since 2LPT performs accurately on very large scales ($k < 0.01 \hMpc$).
The displacement transfer function of 2LPT increases above 1 at $k \approx 0.35 \hMpc$ and then drops sharply.
The increase of 2LPT displacement transfer function is because 2LPT 
over-estimates the displacement power at small scales~\cite[see, e.g.][]{Chan14}. There is a sharp drop of power near the voxel scale because smoothing over voxel scales in our predictions automatically erases power at scales smaller than the voxel size. 

Now we turn to the \dm predictions: both the density and displacement transfer functions of the
\dm differ from 1 by a mere 0.4\% at scale $k \lesssim 0.4 \hMpc$,
and this discrepancy
only increases to 2\% and 4\% for density field and displacement field
respectively, as $k$ increases to the Nyquist frequency around $0.7 \hMpc$.
The stochasticity hovers at approximately $10^{-3}$ and $10^{-2}$ for most scales. In other words, for both the density and displacement fields the correlation coefficient between the \dm predictions and FastPM simulations, all the way down to small scales of $k = 0.7 \hMpc$ is greater than 90\%.
The transfer function and correlation coefficient of the \dm predictions shows that it can reproduce the structure formation of the Universe from large to semi-non-linear scales. \dm significantly outperforms our benchmark model 2LPT in the 2 point function analysis. 
\dm only starts to deviate from the ground truth at fairly small scales. This is not surprising as the deeply nonlinear evolution at these scales is more difficult to simulate accurately and appears to be intractable by current analytical theories\cite{2016arXiv161009321P}.

\begin{figure*}
\centering
\includegraphics[width=2\columnwidth]{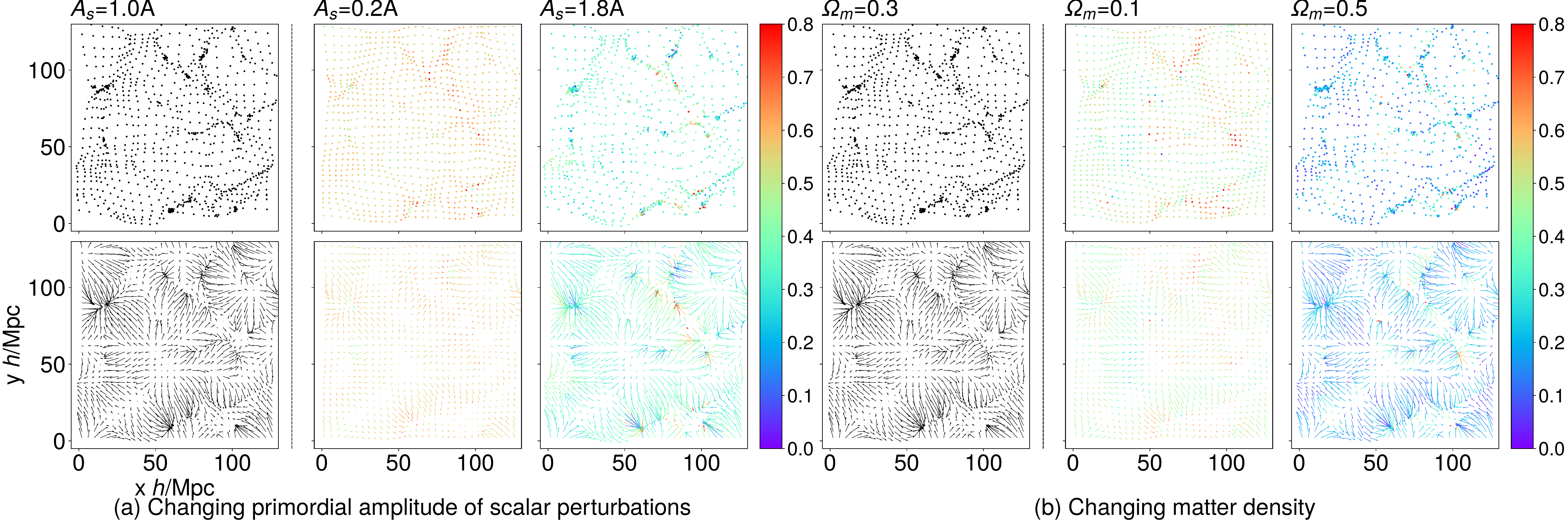}
%\hbox{
%\subfigure[Changing primordial amplitude of scalar perturbations]{
%\includegraphics[width=\columnwidth]{Learning_structure_formation/figures/fig4a.pdf}}
%\subfigure[Changing matter density]{\includegraphics[width=\columnwidth]{Learning_structure_formation/figures/fi%g4b.pdf}}
%}
\caption{
We show the differences of particle distributions and displacement fields when we change the cosmological parameters $A_s$ and $\Omega_m$.\\ 
(\textbf{a}) The errorbar shows the difference of particle distribution (upper panel) and displacement fields (lower panel) between $A_s= A_0$ and the two extremes for $A_s=.2 A_0$(Center) and $A_s=1.8 A_0$ (Right).\\
(\textbf{b}) A similar comparison showing the difference of the particle distributions (upper panel) and displacement fields (lower panel) for smaller and larger values of $\Omega_m \in \{.1,.5\}$ with regard to $\Omega_m = 0.3089$, which was used for training.\\
While the difference for smaller value of $A_s$ ($\Omega_m$) is larger, the displacement for larger $A_s$ ($\Omega_m$) is more non-linear. This non-linearity is due to concentration of mass and makes the prediction more difficult. 
}
\label{fig:generalization}
\end{figure*}

\subsection*{3-Point Correlation Comparison} 
The 3-point correlation function (3PCF) expresses the correlation of the field of interest among 3 locations in the configuration space, which is equivalently defined as bispectrum in Fourier space. Here we concentrate on the 3PCF for computational convenience: 

\begin{equation}
    \zeta(r_1, r_2, \theta) = \langle \delta(\bm{x}) \delta(\bm{x} + \bm{r_1}) \delta(\bm{x} + \bm{r_2}) \rangle.
\end{equation}
where $r_1$=$|\bm{r_1}|$ and $r_2$=$|\bm{r_2}|$. 
Translation invariance guarantees that $\zeta$ is independent of $\bm x$. Rotational symmetry further eliminates all direction dependence except dependence on $\theta$, the angle between $\bm r_1$ and $\bm r_2$. The multipole moments of $\zeta(r_1, r_2, \theta)$, $\zeta_\ell(r_1,r_2) = (2\ell+1) \int d\theta P_\ell(\cos \theta) \zeta(r_1, r_2, \theta)$ where $P_\ell(\cos \theta)$ is the Legendre polynomial of degree $\ell$, can be efficiently estimated with pair counting \cite{Slepian:2015qza}. While the input (computed by ZA) do not contain significant correlations beyond the second order (power spectrum level), we expect \dm to generate densities with a 3PCF that mimics that of ground truth.

We compare the 3PCF calculated from FastPM, 2LPT and \dm by analyzing the 3PCF through its multipole moments $\zeta_\ell(r_1,r_2)$. Figure~\ref{fig:test}(b) shows the ratio of the binned multipole coefficients of the two 3PCF for several triangle configurations, $\bar{\xi_{\ell}}(r_1,r_2)_{pred}/\bar{\xi_{\ell}}(r_1,r_2)_{true}$, where $\bar{\xi_{\ell}}(r_1,r_2)_{pred}$ can be the 3PCF for D$^3$M or 2LPT and $\bar{\xi_{\ell}}(r_1,r_2)_{true}$ is the 3PCF for FastPM. We used 10 radial bins with $\Delta r = 5 \Mpch$. The results are averaged over 10 test simulations and the errorbars are the standard deviation. The ratio shows the 3PCF of D$^3$M is more close to FastPM than 2LPT with smaller errorbars. 
To further quantify our comparison, we calculate the relative 3PCF residual defined by 
\begin{eqnarray}
& &\textnormal{3PCF relative residual}  \nonumber \\
& &= \frac{1}{9 \times N_r}\sum_{\ell=0}^8\sum_{r_1,r_2}\frac{|\zeta_{\ell}(r_1,r_2)_\pred-\zeta_{\ell}(r_1,r_2)_\true|}{|\zeta_{\ell}(r_1,r_2)_\true|}
\end{eqnarray}
where $N_r$ is the number of ($r_1$,$r_2$) bins. The mean relative 3PCF residual of the \dm and 2LPT predictions compared to FastPM are $0.79\%$ and $7.82\%$ respectively. 
%The \dm accuracy on 3PCF is impressive especially given that the input ZA has zero 3 point function. 
The \dm accuracy on 3PCF is also an order of magnitude better than 2LPT, which indicates that the \dm is far better at capturing the non-Gaussian structure formation.

\section*{Generalizing to New Cosmological Parameters} 
So far, we train our model using a ``single'' choice of cosmological parameters $\As= 2.142 \times 10^{-9}$ (hereafter $A_0 = 2.142 \times 10^{-9}) $ and $\Omega_m=0.3089$ ~\cite{Planck15}.
$\As$ is the primordial amplitude of the scalar perturbation from cosmic inflation, and $\Omega_m$ is the fraction of the total energy density that is matter at the present time, and we will call it matter density parameter for short. The true exact value of these parameters are unknown and different choices of these parameters change the large-scale structure of the Universe; see Figure \ref{fig:generalization}.

Here, we report an interesting observation: the \dm trained on a single set of parameters in conjunction with ZA (which depends on $\As$ and $\Omega_m$) as input, can predict the structure formation for widely different choices of $\As$ and $\Omega_m$.
From a computational point of view this suggests a possibility of producing simulations for a diverse range of parameters, with minimal training data.

\subsection*{Varying Primordial Amplitude of Scalar Perturbations $\mathbf{\As}$} 
After training the \dm using $\As=A_0$, we change $\As$ in the input of our test set by nearly one order of magnitude: $A_s = 1.8 A_0$ and $\As = 0.2 A_0$. Again, we use 1000 simulations for analysis of each test case.
The average relative displacement error of \dm remains less than $4\%$ per voxel (compared to $< 3\%$ when train and test data have the same parameters). This is still well below the error for 2LPT, which has relative errors of $15.5\%$ and $6.3\%$ for larger and smaller values of $A_s$ respectively.   

Figure~\ref{fig:A}(a) shows the transfer function and correlation coefficient for both \dm and 2LPT. The \dm performs much better than 2LPT for $A_s = 1.8 A_0$. 
For small $A_s = 0.2 A_0$, 2LPT does a better job than \dm predicting the density transfer function and correlation coefficient at the largest scales, otherwise \dm predictions are more accurate than 2LPT at scales larger than $k=0.08 \hMpc$. 
We observe a similar trend with 3PCF analysis: 
 the 3PCF of \dm predictions are notably better than 2LPT ones for larger $A_s$, compared to smaller $A_s$ where it is only slightly better.
These results confirm our expectation that increasing $A_s$ increases the non-linearity of the structure formation process. While 2LPT can predict fairly well in linear regimes, compared to \dm its performance deteriorates with increased non-linearity. 
It is interesting to note that \dm prediction maintains its advantage despite being trained on data from more linear regimes.

\subsection*{Varying matter density parameter $\mathbf{\Omega_m}$} 
\begin{figure*}
\centering
\includegraphics[width=2\columnwidth]{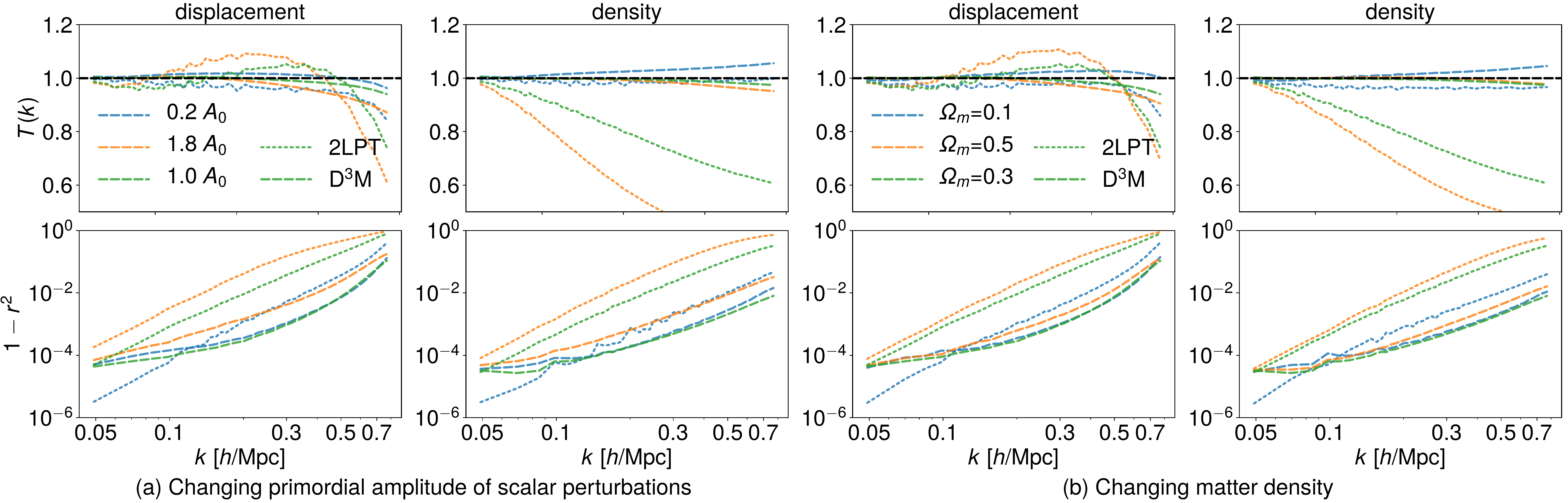}
%\subfigure[Varying Primordial Amplitude of Scalar Perturbations $\As$]
%{\includegraphics[width=.48\columnwidth]{Learning_structure_formation/figures/Pk_dd_A.pdf}
%\includegraphics[width=.48\columnwidth]{Learning_structure_formation/figures/Pk_mm_A.pdf}}
%\subfigure[Varying matter density parameter $\Omega_m$]
%{\includegraphics[width=0.48\columnwidth]{Learning_structure_formation/figures/Pk_dd_Om.pdf}
%\includegraphics[width=0.48\columnwidth]{Learning_structure_formation/figures/Pk_mm_Om.pdf}}
\caption{Similar plots as in Figure.~3(a), except we test the 2 point statistics when we vary the cosmological parameters without changing the training set (which has differnet cosmological parameters) nor the trained model. 
We show predictions from \dm and 2LPT when tested on different (\textbf{a}) $\As$ and, (\textbf{b}) $\Omega_m$. 
We show (top) the transfer function  -- \ie the square root of the ratio of the predicted power-spectrum to the ground truth and (bottom) 1-$r^2$ where $r$ is the correlation coefficient between the predicted fields and the true fields.
\dm prediction outperforms 2LPT prediction at all scales except in the largest scales as the perturbation theory works well in linear regime (large scales).
}
\label{fig:A}
\end{figure*}

We repeat the same experiments, this time changing $\Omega_m$ to $0.5$ and $0.1$, while the model is trained on $\Omega_m = 0.3089$, which is quite far from both of the test sets.  For $\Omega_m = 0.5$ the relative residual displacement errors of the \dm and 2LPT averaged over 1000 simulations are 3.8\% and 15.2\% and for $\Omega_m = 0.1$ they are 2.5\% and 4.3\%. Figures~\ref{fig:A}(c)(d) show the two-point statistics for density field predicted using different values of $\Omega_m$. For $\Omega_m=0.5$, the results show that the \dm outperforms 2LPT at all scales, while for smaller $\Omega_m=0.1$, \dm outperforms 2LPT on smaller scales ($k > 0.1 \hMpc$). 
As for the 3PCF of simulations with different values of $\Omega_m$, the mean relative 3PCF residual of the \dm for $\Omega_m=0.5$ and $\Omega_m=0.1$ are 1.7\% and 1.2\% respectively and for 2LPT they are 7.6\% and 1.7\% respectively. 
The \dm prediction performs better at $\Omega_m=0.5$ than $\Omega_m=0.1$. This is again because the Universe is much more non-linear at $\Omega_m=0.5$ than $\Omega_m=0.1$. The \dm learns more non-linearity than is encoded in the formalism of 2LPT.  

\begin{table}
\centering
\scalebox{0.7}{
\begin{tabular}{ccccccc}
\hline\hline
            type & point-wise & \shortstack{$T(k)$\\$k=0.11$  $^\dagger$} & \shortstack{$r(k)$\\ $k=0.11$} &\shortstack{$T(k)$\\$k=0.50 $}& \shortstack{$r(k)$\\ $k=0.50$} & 3PCF  \\
          \hline
    \textbf{test phase} &  &  & &&  &\\
     2 LPT Density             &  N/A    & 0.96 & 1.00 &0.74& 0.94&0.0782 \\
     \dm Density      &  N/A    & 1.00 & 1.00 &0.99& 1.00&0.0079 \\
     2 LPT Displacement        &  0.093  & 0.96 & 1.00 &1.04& 0.90&N/A  \\
     \dm Displacement &  0.028  & 1.00 & 1.00 &0.99& 1.00&N/A  \\
\hline

    \textbf{$\boldsymbol{\As=1.8 A_0}$}&  &  & &&  &\\
     2LPT Density             &  N/A    & 0.93 & 1.00 & 0.49&0.78&0.243 \\
     \dm Density      &  N/A    & 1.00 & 1.00 & 0.98&1.00&0.039 \\
     2LPT Displacement        &  0.155  & 0.97 & 1.00 & 1.07&0.73&N/A  \\
     \dm Displacement &  0.039  & 1.00 & 1.00 & 0.97&0.99&N/A  \\
     \hline
     
     \textbf{$\boldsymbol{\As=0.2 A_0}$}&  &  & &&  &\\
     2LPT Density             &  N/A    & 0.99 & 1.00 & 0.98&0.99&0.024 \\
     \dm Density      &  N/A    & 1.00 & 1.00 & 1.03&1.00&0.022 \\
     2LPT Displacement        &  0.063  & 0.99 & 1.00 & 0.95&0.98&N/A  \\
     \dm Displacement &  0.036  & 1.00 & 1.00 & 1.01&1.00&N/A  \\
     \hline
     $\boldsymbol{\Omega_m=0.5}$&  &  & &&  &\\
     2LPT Density             &  N/A    & 0.94 & 1.00 & 0.58& 0.87&0.076 \\
     \dm Density      &  N/A    & 1.00 & 1.00 & 1.00& 1.00&0.017 \\
     2LPT Displacement        &  0.152  & 0.97 & 1.00 & 1.10& 0.80&N/A  \\
     \dm Displacement &  0.038  & 1.00 & 1.00 & 0.98& 0.99&N/A  \\
     \hline
     $\boldsymbol{\Omega_m=0.1}$&  &  & &&  &\\
     2LPT Density             &  N/A    & 0.97 & 1.00 &0.96 & 0.99&0.017 \\
     \dm Density      &  N/A    & 0.99 & 1.00 &1.04 & 1.00&0.012 \\
     2LPT Displacement        &  0.043 & 0.97 & 1.00 &0.97 & 0.98& N/A  \\
     \dm Displacement &  0.025 & 0.99 & 1.00 &1.02 & 1.00& N/A  \\
     \hline
\end{tabular}
}\\
$^\dagger$\footnotesize{The unit of $k$ is $\hMpc$. N/A, not applicable}
\vspace{1em}
\caption{A summary of our analysis.}\label{table:sum}
\end{table}

\section*{Conclusions}
To summarize, our deep model \dm can accurately predict the large-scale structure of the Universe as represented by FastPM simulations, at all scales as seen in the summary table in Table.~\ref{table:sum}. 
Furthermore, \dm learns to predict cosmic structure in the non-linear regime more accurately than our benchmark model 2LPT.
Finally, our model generalizes well to test simulations with cosmological parameters ($\As$ and $\Omega_m$) significantly different from the training set.
This suggests that our deep learning model can potentially be deployed for a ranges of simulations beyond the parameter space covered by the training data (Table~\ref{table:sum}). 
Our results demonstrate that the \dm successfully learns the nonlinear mapping from first order perturbation theory to FastPM simulation beyond what higher order perturbation theories currently achieve. 

%We have no reason to believe that replacing FastPM with exact N-body simulations will deteriorate the performance of our method. 
Looking forward, we expect replacing FastPM with exact N-body simulations would improve the performance of our method.
As the complexity of our \dm model is linear in the number of voxels, we expect to be able to further improve our results if we replace the FastPM simulations with higher resolution simulations. 
Our work suggests that deep learning is a practical and accurate alternative to the traditional way of generating approximate simulations of the structure formation of the Universe. 

%%%%%%%%%%%%%%%%%%%%%%%%%%%%%%%%%%%%%%%%%
%%%%%%%Fig%%%%%%%%%%%%%%%%%%%%%%%%%%%%%%%%%%%%%%%%%

%%%%%%%%%%%%%%%%%%%%%%%%%%%%%%%%%%%%%%%%%%%%%%%%%%
\section*{Materials and Methods}

\subsection*{Dataset} 
The full simulation data consists of 10,000 simulations of boxes with ZA and FastPM as input-output pairs, with an effective volume of 20 (Gpc/h)$^3$  ($190\times10^9 $ly$^3$), comparable to the volume of a large spectroscopic sky survey like Dark Energy Spectroscopic Instrument or EUCLID. We split the full simulation data set into 80\%, 10\% and 10\% for training,
validation and test, respectively. We also generated 1000 simulations for 2LPT for each set of tested cosmological parameters. 

\subsection*{Model and Training}
The \dm adopts the U-Net architecture~\cite{Unet} with 15 convolution or deconvolution layers and approximately $8.4 \times 10^6$ trainable parameters. Our \dm generalizes the standard U-Net architecture to work with three-dimensional data \cite{2016arXiv160604797M, 2019MNRAS.482.2861B, 2018arXiv180400816A}. The details of the architecture are described in the following sections and a schematic figure of the architecture is shown in SI Appendix, Figure.~S2. 
In the training phase, we employ the Adam Optimizer \cite{ADAM} with a learning rate of 0.0001, and first and second moment exponential decay rates equal to 0.9 and 0.999, respectively.
We use the Mean-Squared Error as the loss function (See Loss Function) and $L2$ regularization with regularization coefficient 0.0001.

\subsubsection*{Details of the \dm Architecture}
The contracting path follows the typical architecture of a convolution network.
It consists of two blocks, each of which consists of two successive convolutions of stride 1 and a down-sampling convolution with stride 2.
The convolution layers use 3$\times$3$\times$3 filters with a periodic
padding of size 1 (see Padding and Periodic Boundary) on both sides of each dimension. 
%The padding ensures that the first two layers keep the shape of each channel. The following down-sampling layer reduces the shape of the layer by halving each spatial dimension. 
Notice that at each of the two down-sampling steps, we double the number of feature
channels. At the bottom of the \dm, another two successive convolutions with stride 1
and the same periodic padding as above are applied.
The expansive path of our \dm 
is an inverted version of
the contracting path of the network. 
(It includes two repeated applications of
the expansion block, each of which consists of one up-sampling transposed convolution
with stride 1/2 and two successive convolution of stride 1.
The transposed convolution and the convolution are constructed with 3$\times$3$\times$3
filters.)

We take special care in the padding and cropping procedure to preserve the
shifting and rotation symmetry in the up-sampling layer in expansive path. Before the transposed convolution we apply a periodic padding of length 1 on the
right, down and back sides of the box ({\tt padding=(0,1,0,1,0,1)} in pytorch), and after the transposed convolution, we
discard one column on the left, up and front sides of the box and two columns
on the right, down and back sides ({\tt crop=(1,2,1,2,1,2)}).

A special feature of the \dm is the concatenation procedure, where the up-sampling layer halves the feature channels and then concatenates
them with the corresponding feature channels on the contracting path, doubling the number of feature channels.

The expansive building block then follows a 1$\times$1$\times$1 convolution without padding, which converts the 64 features to the the final 3-D displacement field. % ($\Psi_x$,$\Psi_y$,$\Psi_z$).
All convolutions in the network except the last one are followed by a
rectified linear unit activation and batch normalization (BN). 

\subsubsection*{Padding and Periodic Boundary} 
It is common to use constant or reflective padding in deep models for image processing. However, these approaches are not suitable for our setting.
The physical model we are learning is constructed on a spatial volume with a periodic
boundary condition. This is sometimes also referred to as a torus geometry, where the boundaries of the simulation box are topologically connected -- that is $x_{i+L} = x_{i}$
where $i$ is the index of the spatial location, and $L$ is the periodicity (size of box).
Constant or reflective padding strategies break the connection between the physically nearby points separated across the box, which not only loses information but also introduces noise during the convolution, further aggravated with an increased number of layers. 

We find that the periodic padding strategy significantly improves the
performance and expedites the convergence of our model, comparing to the same
network using a constant padding strategy.
This is not surprising, as one expects it is easier to train a model that can
explain the data than to train a model that does not.

\subsubsection*{Loss Function}
We train the \dm to minimize the mean square error
on particle displacements 
\begin{equation}
    \mathcal{L} = \frac{1}{N}\sum_i (\hat{\vPsi}_{i} - \vPsi_{t,i})^2,
\end{equation}
where $i$ labels the particles and the N is the total number of particles.
This loss function is proportional to the integrated squared error,
and using a Fourier transform and Parseval’s theorem it can be rewritten as
\begin{align}
    &\int (\hat{\vPsi} - \vPsi_t)^2 \d^3\vq
    = \int \bigl|\hat{\vPsi} - \vPsi_t\bigr|^2 \d^3\vk =
    \nonumber\\
    & \int \d^3\vk \Biggl( \bigl|\vPsi_t\bigr|^2 (1- T)^2
     + 2\bigl|\hat{\vPsi}\bigr|\bigl| \vPsi_t\bigr| (1-r) \Biggr)
\label{Parseval}
\end{align}
where $\vq$ is the Lagrangian space position, and $\vk$ its corresponding wavevector.
$T$ is the transfer function defined in Eq.~\ref{transfer},
and $r$ is the correlation coefficient defined in Eq.~\ref{corrcoeff}, which characterize the similarity between the predicted and the true fields, in amplitude and phase respectively.
Eq.~\ref{Parseval} shows that our simple loss function jointly captures both of these measures:
as $T$ and $r$ approach 1, the loss function approaches 0.

\subsubsection*{Data availability}
The source code of our implementation is available at \url{https://github.com/siyucosmo/ML-Recon}. The code to generate the training data is also available at \url{https://github.com/rainwoodman/fastpm}.
\showmatmethods{}

\acknow{ We thank Angus Beane, Peter Braam, Gabriella Contardo, David Hogg, Laurence Levasseur, Pascal Ripoche, Zack Slepian and David Spergel for useful suggestions and comments, Angus Beane for comments on the paper, Nick Carriero for help on Center for Computational Astrophysics (CCA) computing clusters. The work is funded partially by Simons Foundation.
The FastPM simulations are generated on the computer cluster Edison at the National Energy Research Scientific Computing Center (NERSC), a U.S. Department of Energy Office of Science User Facility operated under Contract No. DE-AC02-05CH11231. The training of neural network model is performed on the CCA computing facility and the Carnegie Mellon University AutonLab computing facility. The open source software toolkit \texttt{nbodykit} \cite{nbodykit} is employed for the clustering analysis. YL acknowledges support from the Berkeley Center for Cosmological Physics and the Kavli Institute for the Physics and Mathematics of the Universe, established by World Premier International Research Center Initiative (WPI) of the MEXT, Japan. S.Ho thanks NASA for their support in grant number: NASA grant 15-WFIRST15-0008 and NASA Research Opportunities in Space and Earth Sciences grant 12-EUCLID12-0004, and Simons Foundation.}

\showacknow{} % Display the acknowledgments section

% Bibliography
\section*{References}
%\bibliography{ref}

\end{document}